# Bandstructure and Mobility Variations in p-type Silicon Nanowires under Electrostatic Gate Field

Neophytos Neophytou, Oskar Baumgartner, Zlatan Stanojevic, and Hans Kosina

Institute for Microelectronics, TU Wien, Gußhausstraße 27-29/E360, A-1040 Wien, Austria

e-mail: neophytou@iue.tuwien.ac.at

## Abstract

The $sp^3d^5s^*$-spin-orbit-coupled atomistic tight-binding (TB) model is used for the electronic structure calculation of Si nanowires (NWs), self consistently coupled to a 2D Poisson equation, solved in the cross section of the NW. Upon convergence, the linearized Boltzmann transport theory is employed for the mobility calculation, including carrier scattering by phonons and surface roughness. As the channel is driven into inversion, for [111] and [110] NW devices of diameters D>10nm the curvature of the bandstructure increases and the hole effective mass becomes lighter, resulting in a ~50% mobility increase. Such improvement is large enough to compensate for the detrimental effect of surface roughness scattering. The effect is very similar to the bandstructure variations and mobility improvement observed under geometric confinement, however, in this case confinement is caused by electrostatic gating. We provide explanations for this behavior based on features of the heavy-hole band. This effect could be exploited in the design of p-type NW devices. We note, finally, that the "apparent" mobility of low dimensional short channel transistors is always lower than the intrinsic channel diffusive mobility due to the detrimental influence of the so called "ballistic" mobility.

# I. Introduction

Silicon nanowires (NWs) and narrow multi-gate or gate-all-around devices have recently attracted significant attention as candidates for the next generation transistor devices. Such channels offer the possibility of enhanced electrostatic control and of quasi-ballistic transport [1]. Ultra-scaled NW devices of channel diameters down to D=3nm and lengths as short as 15nm with excellent performance have already been demonstrated by various experimental groups [2, 3, 4, 5, 6, 7, 8]. Furthermore, large mobility and on-current densities in NWs compared to bulk have been observed experimentally [9, 10]. Low-dimensional channels offer additional degrees of freedom in engineering their properties: i) the length scale of the cross section, ii) the transport orientation, iii) the orientation of the confining surfaces [11, 12]. The transport effective mass, carrier velocity, mean free path, and mobility can significantly vary with geometry and strongly influence the performance of devices.

Several simulation works have recently examined the performance of ultra-narrow channels addressing a variety of the above mentioned factors. In those works the electronic structure was calculated using approaches varying from continuum effective mass descriptions [13, 14], to k·p [15, 16, 17] and atomistic tight-binding (TB) [11, 12, 18, 19, 20, 21, 22], or even DFT [23]. Transport methodologies from semiclassical to fully quantum mechanical were also utilized. Accurate description of the electronic structure is especially important for p-type channels, where strong geometrical confinement, spin-orbit interactions and band coupling can result in warping and non-trivial features of the electronic structure [12, 24]. For ultra-narrow p-type NWs, studies have shown that the carrier velocities and mobility largely increase as the diameter is reduced down to D=3nm [10, 21, 25]. This was a result of bandstructure changes upon geometrical confinement, and was observed for [111] and [110] NWs.

Most atomistic studies, however, are limited computationally by the cross section of the channel, and usually NWs of diameters below 5nm are considered. In this work, we extend the simulation domain to D=12nm (~5500 atoms), still using the atomistic TB method for the electronic structure. We show that bandstructure effects and band warping are important even for p-type NW transistors with larger cross section. The electrostatic confinement caused by the gate field at inversion conditions can influence the bandstructure in a similar manner as geometrical confinement does. The effective mass of the bands becomes bias dependent, and large increases in the carrier mobility (of the order of ~50%) can be observed in [110] and [111] NWs as the channel is driven into inversion. We provide explanations of this based on features of the heavy-hole valence band of Si. We finally distinguish the influence of this intrinsic, diffusive mobility from the total "apparent" mobility that would be measured from experiments in ultra-scaled channels, and show that the latter will always be lower due to the detrimental effect of the so called "ballistic" mobility.

This paper is organized as follows: In part II we describe the self consistent calculation of the TB electronic structures and their coupling to linearized Boltzmann transport theory. In part III we show how the bandstructure of the p-type NWs is affected under electrostatic inversion for NWs in the [100], [110] and [111] transport orientations, and study the influence on the mobility. In part IV we discuss and present explanations for these observations and part V concludes the work.

## II. Approach

To obtain the bandstructure of the NWs we use the nearest neighbor $sp^3d^5s^*$ tight-binding model [11, 26, 27, 28, 29], which captures all the necessary band features, and in addition, is robust enough to computationally handle larger NW cross sections as compared to *ab-initio* methods. As an indication, the unit cells of the NWs considered in this study contain up to ~5500 atoms. The computation time needed for the calculation of the self consistent bandstructure of these NWs can take several days on a single CPU. Each atom in the NW unit cell is described by 20 orbitals including spin-orbit-coupling. The model itself and the parameterization used [26] have been extensively calibrated to various experimental data with excellent agreement without additional material parameter adjustments [30, 31, 32, 33]. We consider p-type silicon NWs of D=12nm in three different transport orientations [100], [110], and [111].

The complete computational model is described in Fig. 1 [34]. There are four steps in the computation. i) The first step is the calculation of the electronic structure of the NW channel using the TB model. The Schodinger equation is solved only in the Si channel, whereas the oxide is only included in the electrostatics of the device. The TB model assumes hydrogen passivated edge atoms using an effective passivation scheme described in Ref. [29]. Effectively, this scheme places hard wall boundary conditions on the edge atoms of the channel, while removing all dangling bonds that remain from the unsaturated atomic bonds. This mimics Hydrogen passivation, although there are no actual Hydrogen atoms used in the Hamiltonian construction. ii) The second step involves the calculation of the charge density using a semiclassical model. The model used is commonly referred to as the "top-of-the-barrier" model [35], and has traditionally been used in calculating the ballistic performance of transistor devices. In this model the positive and negative k-states of the bandstructure are filled according to the Fermi levels of the source and the drain respectively. In our case, since we calculate low-field mobility using linearized Boltzmann, we only have one Fermi level in the entire device which fills both the positive and negative going states equivalently. The position of the Fermi level determines the threshold voltage and the off-state of the transistor, introducing an effective channel doping (although we do not include actual dopant atoms in the channel). We chose the Fermi level such that all NWs have the same threshold voltage, around $V_G$=-0.2V. This is shown in Fig. 2. Figure 2a shows the position of the valence band edge with respect to the Fermi level $\eta_F=E_V-E_F$. As the gate bias increases (more negative), the valence band is pushed up above the Fermi level, in a similar amplitude for all three NW orientations. Accordingly, the charge density in the cross section of the NWs increases as shown in Fig. 2b, and again, does so very similarly for the three NW orientations. iii) The third step is the solution of a 2D Poisson equation in the cross section of the NW. A gate all around geometry is used, with 1.2nm $SiO_2$ as the gate insulator. Since the position of the Fermi level determines the threshold voltage, in the model we do not need to consider a specific value for the workfunction difference between the gate metal and the oxide. This is indirectly included in the choice of the position of the Fermi level to achieve a specific threshold voltage. When solving the Poisson equation we also include the carrier distribution in the channel, which is determined by the coefficients of the eigenvectors of the various k-states. For example, Fig. 3a and 3b show the charge density distribution in the cross section of the [111] NW under low gate bias conditions ($V_G$=-0.2V) and inversion ($V_G$=-0.8V), respectively, at $V_D$=0V. In the first case the charge is distributed uniformly in the cross section of the NW, whereas in the second case, surface inversion is observed. As we will explain below, this electrostatic confinement has strong effects on the bandstructure of the NW. These first three steps are solved self-consistently since the bandstructure is a function of the potential profile in the channel. More details of the model can be found in our previous works [11, 12, 24]. iv) Once self-consistency is achieved, we use the linearized Boltzmann transport formalism to compute the low-field mobility of the NW channel. We perform the above procedure for a series of gate biases, driving the channel from depletion to inversion and further on to strong inversion conditions.

We describe the linearized Boltzmann transport approach we describe in detail in Ref. [36]. In brief, the electrical low-field conductivity ($\sigma$) follows from the linearized Boltzmann equation as

$$\sigma = q_0^2 \int_{-\infty}^{E_0} dE \left( -\frac{\partial f(E)}{\partial E} \right) \Xi(E). \tag{1}$$

The energy *(E)* integration over the derivative of the Fermi distribution $f(E)$ is performed though all energies up to the valence band edge $E_0$. $\Xi(E)$ is the transport distribution function (TD) defined as [37, 38]:

$$\begin{aligned}\Xi(E) &= \frac{1}{A} \sum_{k_x,n} v_n^2(k_x) \tau_n(k_x) \delta(E - E_n(k_x)) \\ &= \frac{1}{A} \sum_n v_n^2(E) \tau_n(E) g_{1D}^n(E).\end{aligned} \tag{2}$$

Here $v_n(E) = \frac{1}{\hbar}\frac{\partial E_n}{\partial k_x}$ is the group velocity of a hole in subband $n$ with dispersion $E_n(k_x)$, $\tau_n(k_x)$ is the momentum relaxation time for a carrier with wavenumber $k_x$ in subband $n$,

$$g_{1D}^n(E) = \frac{1}{2\pi\hbar}\frac{1}{|v_n(E)|} \quad (3)$$

is the density of states for the 1D subbands (per spin), and $A$ is the cross sectional area of the NW. The mobility ($\mu$) is defined as:

$$\mu = \frac{\sigma}{q_0 \mathrm{p}}, \quad (4)$$

where p is the carrier concentration in the channel and $q_0$ is the elementary charge.

For the calculation of the relaxation times we use Fermi's golden rule. We include acoustic phonons, optical phonons and surface roughness scattering (SRS). We use bulk phonons and bulk phonon deformation potential values from Ref. [39]. The NW diameters used are D=12nm, which are large enough such that the effect of phonon confinement is insignificant [21, 40, 41, 42, 43]. The details of the formalism for the extraction of the phonon scattering rates from the atomistic bandstructures are explained in detail in Ref. [36]. For SRS we derive the scattering matrix element and scattering rates from the strength of the electric field in the channel [40]. The transition rate is given by:

$$S_{n,m}(k_x, k_x') = \frac{2\pi}{\hbar}\left|\langle F_f | E_{eff}(r) | F_i \rangle\right|^2 \frac{\Delta_{rms}^2 L_C}{(2+q_x^2 L_C^2)} \delta(E_m(k_x') - E_n(k_x)) \quad (5)$$

Above, $\delta(\cdot)$ is the Dirac-delta function which denotes energy conservation, $F_{f,i}$ are the final/initial bound states in the transverse plane (NW cross section), $E_{eff}$ is the radial gate-induced electric field in the NW cross section, and $q_x = k_x - k_x'$. We assume a 1D exponential autocorrelation function [44] for the roughness where $\Delta_{rms}$ = 0.48nm is the roughness amplitude and $L_C$ =1.3nm is the roughness correlation length. Note that this is a different treatment of SRS than the one we describe in Ref. [36]. There we employ the linearized Boltzmann formalism in NWs with flat electrostatic potential in the cross section, and consider the band edge shift as the dominant influence of the SRS. That mechanism is the dominant for ultra-narrow NWs, of diameters D<8nm, and is responsible for the $\mu \sim D^6$ mobility behavior observed in ultra thin-layers and NWs is [14, 45]. For the diameters of interest in this new work, i.e. D=12nm, the bandstructure approaches bulk, and that effect is minimal, and therefore not included here. At first order, it could be included as part of a somewhat larger $\Delta_{rms}$ compared to the one we have used. In such way it would only lead to a small quantitative change in the results, but does not affect our conclusions.

## III. Results

We have shown in previous works that the electronic structure of ultra-narrow NWs is sensitive not only to orientation, but also to diameter [24, 25]. Below we will show that similar bandstructure changes are observed not only under geometrical confinement, but also under electrostatic confinement i.e. by gating. Figure 4 describes these changes for p-type NWs of D=12nm in the three transport orientations [100], [110], and [111]. Row-wise, Fig. 4 shows the electronic structures for the three different orientations. Column-wise the figure depicts: i) Bandstructures under flat band conditions (depletion) and D=12nm. ii) Bandstructures under inversion conditions with $V_G$=-0.8V. The position of the Fermi level is indicated. iii) Bandstructures under strong inversion conditions with $V_G$=-1.4V. iv) The bandstructures of NWs with D=3nm (geometrical confinement) under flat potential. The main observations regarding the shape of the bandstructure can be summarized from these figures as follows: i) In the [100] transport orientation (first row), the curvature of the bandstructure does not change significantly, neither by electrostatic, nor by geometrical confinement. ii) In the [110] and [111] transport orientations, under electrostatic confinement the bands become lighter (Fig. 4f-g and Fig. 4j-k), a similar behavior that is observed under geometrical confinement (Fig. 4h, and Fig. 4l). It is evident, therefore, that the effective mass of p-type NW channels is also bias dependent.

This bandstructure variation with gate bias influences the low-field mobility of the NWs. Figure 5a shows the mobility versus gate bias for the NWs in the three different orientations. The mobility is anisotropic with the [111] performing better than the [110] and both better than the [100] NWs. Previous work have shown that the effective mass of the [111] and secondly of the [110] NW is lighter than that of the [100] NW, and this provides an advantage in their transport capabilities [21, 24, 25]. The interesting point in this work, however, is that in the first two cases, the mobility undergoes an increase of ~50% as the channel is driven into inversion. A mobility peak is reached around $V_G$=-0.7V. The increase is strong enough to compensate for the detrimental effect of SRS, as shown by the dotted lines in Fig. 5a which depict simulation results including phonons and SRS. This behavior is explained by the fact that the bands become lighter as shown in Fig. 4f, 4g, 4j, and 4k, which increases the carrier velocities and the mobility. As the gate bias increases even further, the mobility drops back to the low bias values. As observed in Fig. 4g and 4k, the Fermi level at strong inversion is located in regions of heavier subbands with large degeneracy, which increases scattering and causes mobility reduction, as has been also reported for thin p-type layers [46].

We note that the increase observed in the mobility is a consequence of the increase in the curvature of the bands with electrostatic confinement. At smaller NW diameters, the band curvature is already increased because of geometrical confinement. The gate bias induced mobility increase for narrower NW diameters is smaller than in the larger NW diameters, although the absolute value of the mobility is higher as shown in Ref. [25]. This is shown in Fig. 5b, which plots the percentage of the bias induced mobility increase (between the low bias mobility and the highest mobility) for NWs of different diameters. Indeed, the NWs with the largest diameter exhibit the largest gate-induced mobility increase. The mobility increase starts to appear at diameters above D~6nm. This should not be confused with the absolute mobility value of NWs of different diameters. For small diameters, the phonon limited mobility in the [110] and [111] NWs is actually largely increased because the curvature the bandstructure is increased by confinement. For example, the phonon-limited mobility of the D=3nm [111] NW can reach ~2000cm$^2$/V-s [25]. This is much higher than that of the D=12nm NW.

The gate bias does not cause any additional increase in the mobility of the narrower NWs because the curvature of the bands is already increased by structural confinement. However, it causes a 50% increase in the mobility of the larger diameter NWs.

The monotonic trend in the mobility increase with diameter should of course saturate once the diameter is large enough and the influence of the structural confinement is minimized. We expect that to saturate soon after diameter D=12nm since the bandstructure of the D=12nm NW already approaches bulk. This is difficult to verify by simulation because the current structures already include 5500 atoms, and it is computationally challenging to extend the simulation domain beyond that. The observed mobility increase is related to surface inversion. The bandstructure in a large diameter NW is very similar to the bulk heavy-hole bandstructure. At high -$V_G$, the carriers are confined on the surface. The resulting bands can have much lighter masses depending on which surface they reside on [47]. Therefore, once the bands can be transformed from bulk-like to surface-like, we will observe an increase in mobility. However, this is the case only for the (110), and (112) surfaces. The (100) surface, also the one used in traditional planar MOSFET devices, does not offer this advantage and what shown in Fig. 5b is not observed.

For short channels, semi-ballistic transistors, however, the diffusive mobility definition loses its validity. The measured mobility in a classical MOSFET is better explained by the combination of the diffusive and the so-called "ballistic" mobility defined as $\mu_B = L_G / C_{OX}(V_G - V_T) R_{CH}$ [48]. In this expression $L_G$ is the channel length, $C_{OX}$ is the oxide capacitance, and $V_T$ is the threshold voltage. The "ballistic" mobility of this depends on the channel resistance $R_{CH}$, extracted from the slope of the $I_D$-$V_D$ curve at high $V_G$. In Ref. [48] we have indicated that even under zero series resistance, in low dimensional channels the "ballistic" mobility is limited to a few hundred cm$^2$/Vs. This is a consequence of the quantum resistance $h/q_0^2$ associated with each quantum mechanical mode that the carriers occupy. Here, $h$ is Planck's constant. For 3D channels, this quantity is not important since when a lot of modes contribute to transport its value is minimized. For 1D channels, however, the value of the quantum resistance is large enough to affect transport. The slope of the $I_D$-$V_D$ for low $V_D$ and high $V_G$ is determined by the series combination of the quantum resistance and the channel resistance. As the channel is shortened, the channel resistance is reduced (and completely eliminated in the ballistic case), and the only resistance that remains is the quantum resistance. Because of this low dimensional effect, when extracting the mobility for 1D channel transistors using the usual method from the slope of the linear region of the $I_D$-$V_D$, one will extract a smaller value compared to the one expected from the intrinsic part of the channel alone. The computed ballistic mobility $\mu_B$ versus -$V_G$ for the NWs in the three orientations is shown in Fig. 6. Its magnitude is lower than the diffusive mobility in Fig. 5 (also shown by the dashed line in Fig. 6 for the [111] NW case), and its trend is different.

The "ballistic" mobility will be more relevant for short channels, whereas the diffusive mobility is important for long channels. This is demonstrated in Fig. 7, which shows the total mobility $\mu_{TOT}$ (so-called "apparent" mobility) at inversion conditions ($V_G$=-0.6V) versus channel length $L_G$. This quantity is calculated from the diffusive and ballistic mobilities using Matthiesen's rule as $1/\mu_{TOT} = 1/\mu_D + 1/\mu_B$ [19, 49, 50, 51]. For longer channels, the mobility approaches the diffusive mobility values (peaks of Fig. 5), whereas it approaches the "ballistic" mobility values as the channels shortens (Fig. 6 values). Nevertheless, even in the case of scaled channels, mechanisms that provide diffusive mobility improvements can improve the ballisticity of the device by achieving ballistic behavior for longer channel lengths, and improve the speed of transistors.

Another point worth mentioning is that in addition to enhanced surface roughness scattering, nanowire channels could suffer from various other scattering mechanisms that are enhanced under strong confinement, or in ultra-narrow channels. An important mobility detrimental mechanism is the effect of trap densities, whose concentration is larger in NWs than in planar structures [52, 53]. Mechanisms that improve mobility by confinement, however, could offset these effects and allow NWs with high performance. We mention that the density of traps is significantly larger for cylindrical NWs in which the surface orientation varies continuously, compared to rectangular NWs as discussed in references [52, 53]. Although we simulate cylindrical NWs, in our simulations we have not considered such effects because we focus on the effect of bandstructure and how it is affected by potential variations. We expect similar observations in the case of rectangular NWs, for which the measured mobility could be higher than that of the cylindrical NWs due to the reduced amount of defects.

## IV. Discussion

In this section we provide explanations as of why the curvature of the bandstructure changes with electrostatic confinement in the [110] and [111] NWs. It turns out that at first order, the warped shape of the heavy-hole band as shown in Fig. 8 can explain this behavior. The heavy-hole band has a warped structure consisting of pairs of "wings" that project differently on each surface [17]. The four "wings" can be thought of as a superposition of two ellipsoids. Each pair of "wings" reacts differently to geometrical confinement [25], strain [17], and as we will show here, electrostatic confinement. Similar effects have been observed not only for holes, but also for electrons in warped conduction bands under either electrostatic confinement [11], or strain [54, 55, 56].

Figure 8a shows the (100) heavy-hole energy surface of the valence band. There are four "wings" (or a pair of "ellipsoids" placed perpendicular to each other) in this case. For NWs in the [100] transport direction, this is the relevant in-plane energy surface. Under cylindrically symmetric electrostatic confinement the four "wings" behave equivalently as indicated in Fig. 8b. For [110] NWs, however, the situation is different as shown in Fig. 8c. In this case, the transport direction is the [110] direction, inclined at 45º, whereas the confinement direction is along the [1-10] direction, at 135º. From here, it is evident that two "ellipsoids" (red and blue) will behave differently. The one having its long axis along the transport direction (red) has a large transport mass, but a light confinement mass. Under electrostatic confinement, this "ellipsoid" will shift higher in energy than the other ellipsoid (blue), which has a larger confinement mass. The second

"ellipsoid" (blue), however, has a light transport mass. As the confinement increases, the second "ellipsoid" has more contribution, and the electronic structure shows a lighter mass, as observed in Fig. 4. In reality there is a stronger band coupling and band warping under confinement conditions, but at first order, the shape of the heavy-hole band can qualitatively explain the bandstructure changes observed. It is also worth mentioning that the other confinement direction is out of the plane along the equivalent [100] direction. The confinement caused by this (010) surface is equivalent for both "ellipsoids" and does not alter the bandstructure. The confinement along the [1-10] direction is therefore the dominant factor for the bandstructure behavior observed in Fig. 4, whereas [100] confinement does not have a strong influence. Indeed, the benefit of the [110] versus [100] confinement for holes has been demonstrated by simulations [57] and by experimental data in thin layers and MOSFET devices of different crystallographic orientations [58, 59, 60].

The situation for the [111] NW is explained in similar terms in Fig. 8d-f. Figure 8d shows the (112) surface, which includes the [111] and [1-10] directions. As shown in Fig. 8e, there are again two "ellipsoids" perpendicular to each other. Similar to Fig. 8c, they are not equivalent under transport in the [111] direction and confinement in the [1-10] direction. Under confinement, the "ellipsoid" with the longitudinal axis along [111] (red), is weakly confined and will be raised in energy compared to the ellipsoid which has its longitudinal axis in the perpendicular [1-10] direction (blue). The latter ellipsoid (blue) has a light transport mass, which makes the dispersion of the [111] NW light as well, as indicated in Fig. 4j-k. We note that in the case of the [111] NW, all confinement orientations have a similar influence on the confinement of the NW dispersion. This is in contrast to the [110] NW, where only the [1-10] confinement direction affects the dispersion curvature. This is shown for the (111) heavy-hole energy surface in Fig. 8f. In the orientations [11-2] or [1-10] the energy surface is more uniform compared to the (100) surface, although warping still can be observed.

## V. Conclusion

The sp$^3$d$^5$s$^*$ atomistic tight-binding model is self-consistently coupled to Poisson equation and the linearized Boltzmann transport formalism for the calculation of mobility in p-type silicon NWs. We show that the phonon-limited low-field mobility in p-type [110] and [111] NWs has a strong gate bias dependence and can increase by ~50% as the channel is driven into inversion. This increase is observed for NWs with diameter of D=12nm and is accounted on bandstructure changes under electrostatic confinement. For smaller diameters the bias-induced mobility increase is reduced because the bands already become lighter by geometrical confinement. We show that at first order, the bandstructure changes with electrostatic confinement can be explained from features of the heavy-hole valence band. The mobility results are relevant for diffusive devices, whereas for short channels, close to ballistic transport, the total mobility ("apparent" mobility) is influenced by the "ballistic" mobility, and becomes lower than its diffusive component. The large variations in mobility versus gate voltage are then partially smeared out. Nevertheless, larger diffusive mobility corresponds to large mean free paths for scattering and larger channel ballisticity, which improves performance.

## Acknowledgements

This work was supported by the Austrian Climate and Energy Fund, contract No. 825467.

Figure 1:

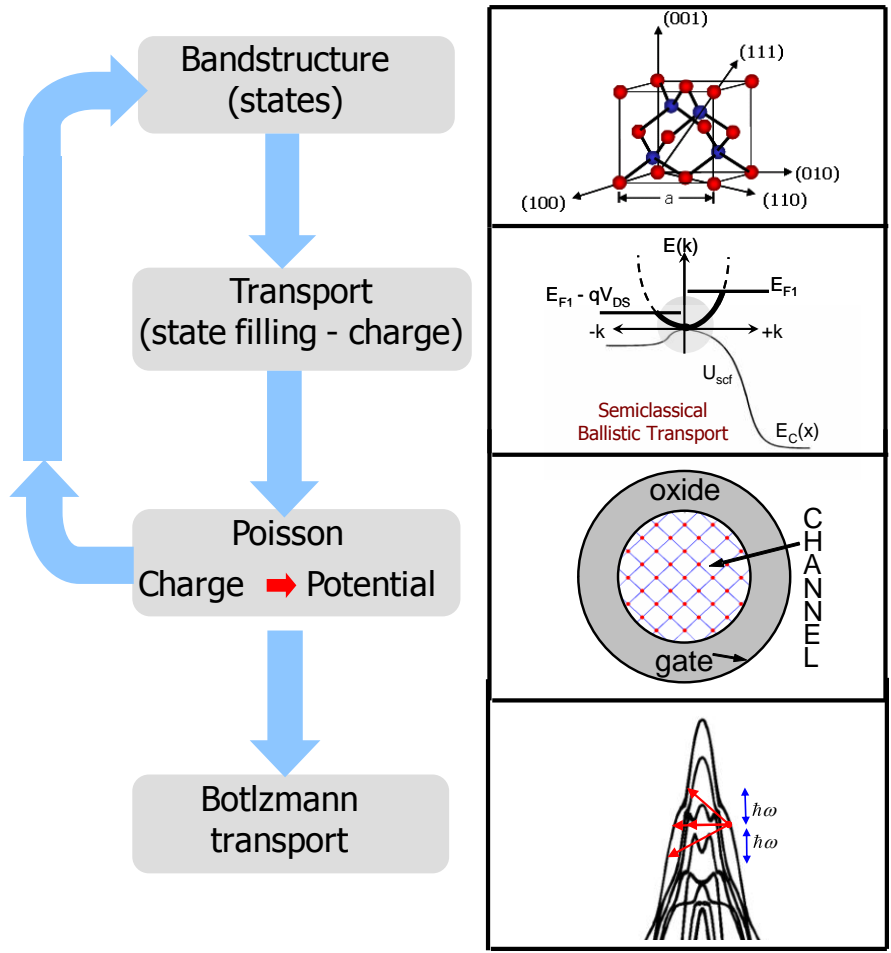

Figure 1 caption:

Simulation procedure steps (from top to bottom). The NW bandstructure is calculated using the $sp^3d^5s^*$ TB model. (b) A semiclassical ballistic model is used to calculate the charge distribution in the NW. (c) The charge is self-consistently coupled to a 2D Poisson equation for the electrostatic potential in the cross section of the wire. From here, ballistic characteristics can be extracted. (d) Upon convergence (and at $V_D$=0V), Boltzmann transport theory is used for mobility calculations. (Examples of relevant valence band scattering mechanisms are shown.)

Figure 2:

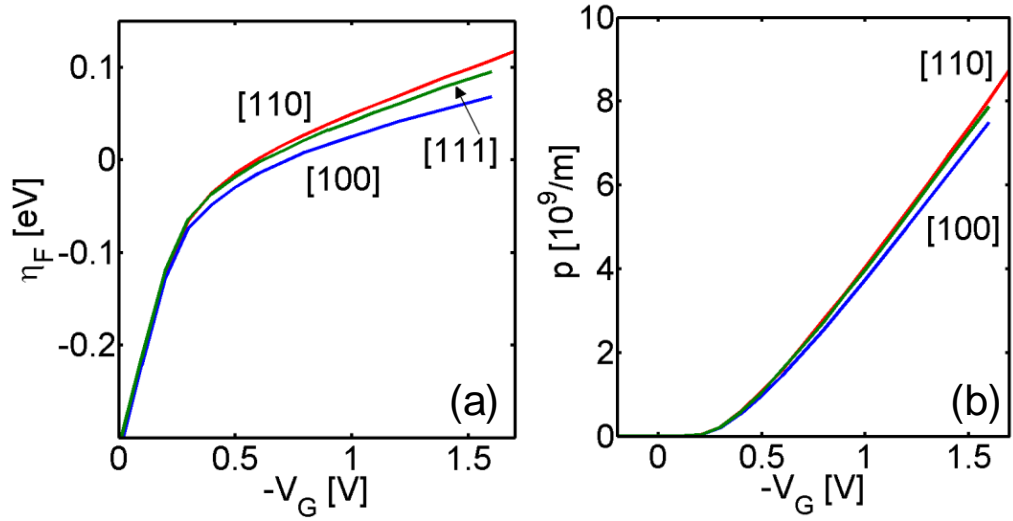

Figure 2 caption:

(a) The difference of the valence band from the Fermi level $\eta_F = E_V - E_F$ versus gate bias. (b) The charge density in the NWs versus gate bias. Quantities for NWs of D=12nm in the [100] (blue), [110] (red) and [111] (green) transport orientations are indicated.

Figure 3:

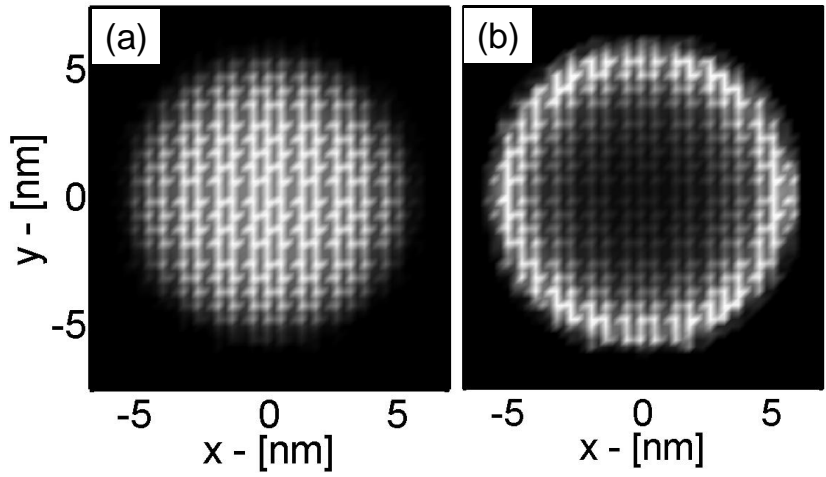

Figure 3 caption:

The hole distribution in the cross section of the [111] NW of D=12nm, at (a) off-state, and (b) on-state ($V_G$=-0.8V).

Figure 4:

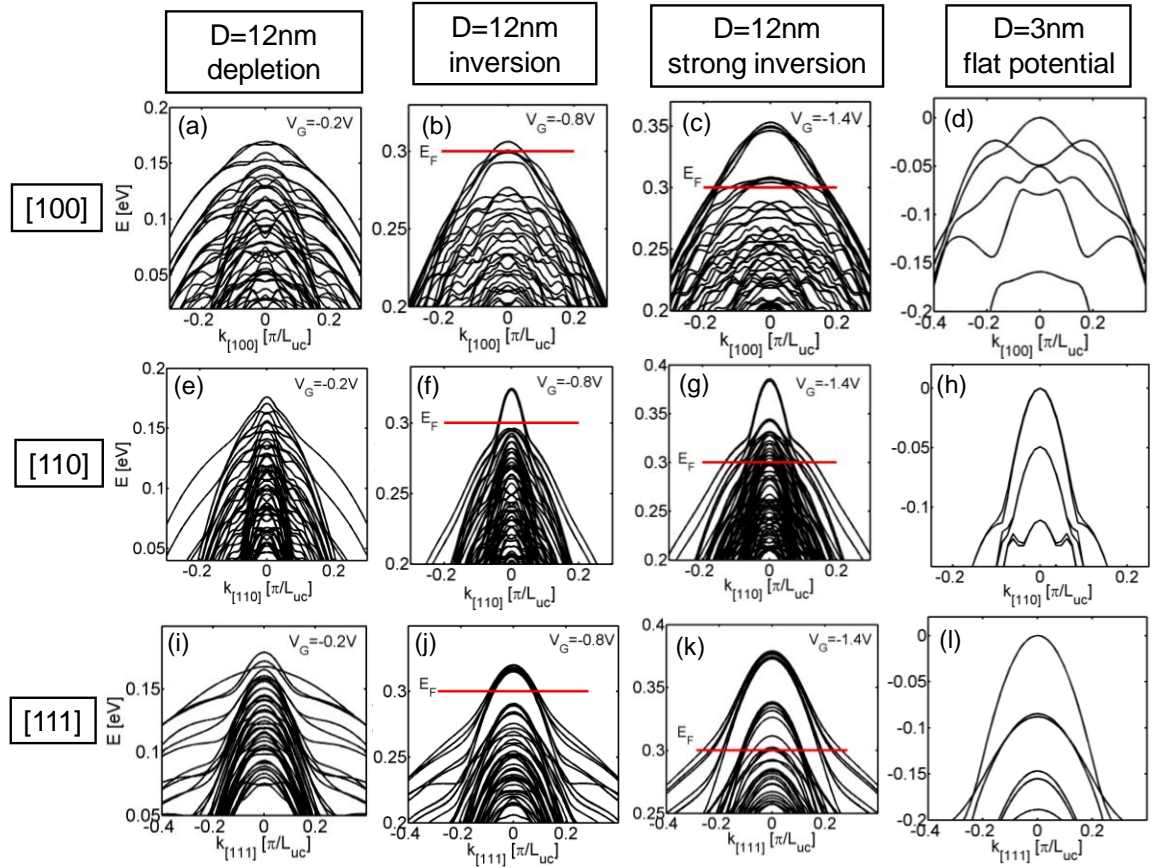

Figure 4 caption:

Dispersions of NWs under different confinement conditions. Row-wise: [100], [110] and [111] transport orientations, respectively. Column-wise: i) D=12nm, $V_G$=-0.2V (depletion), ii) D=12nm, $V_G$=-0.8V (inversion), iii) D=12nm, $V_G$=-1.4V (strong inversion), iv) D=3nm, flat potential.

Figure 5:

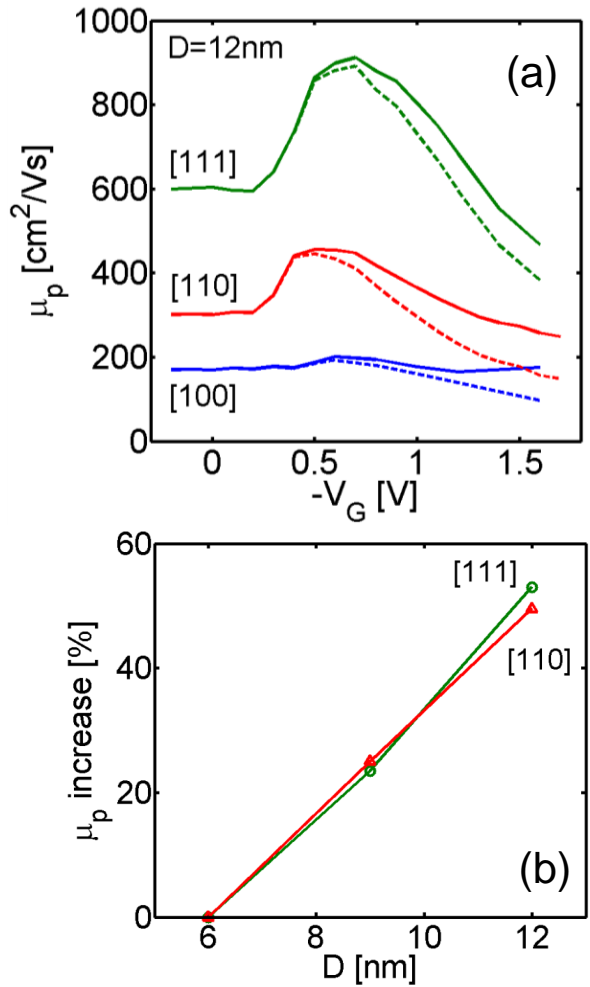

Figure 5 caption:

(a) Low-field hole mobility for NWs of D=12nm in the [100], [110], and [111] transport orientations versus gate bias (-$V_G$). Solid lines: Phonon-limited mobility. Dashed lines: Phonon plus surface roughness scattering limited mobility. (b) The percentage of gate-induced mobility increase in the [110] and [111] NWs versus diameter. The circle-green line is for the [111] NW and the triangle-red for the [110] NW.

Figure 6:

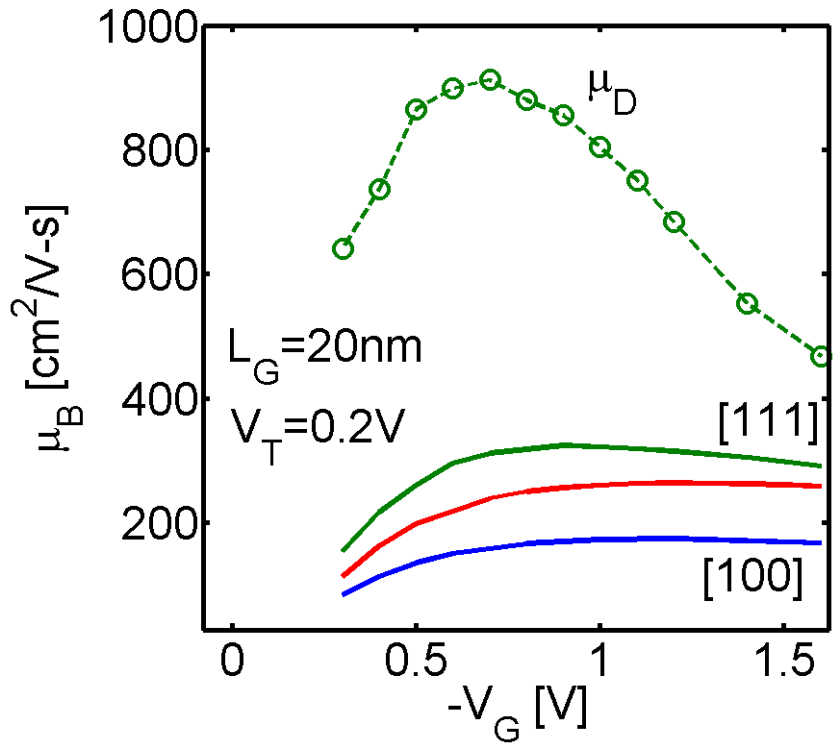

## Figure 6 caption:

Solid lines: Hole "ballistic" mobility for NWs of D=12nm in the [100] (blue), [110] (red), and [111] (green) transport orientations versus (-$V_G$). A device with channel length $L_G$=20nm is assumed. Dashed line: The diffusive mobility of the [111] NW case, same as in Fig. 5.

Figure 7:

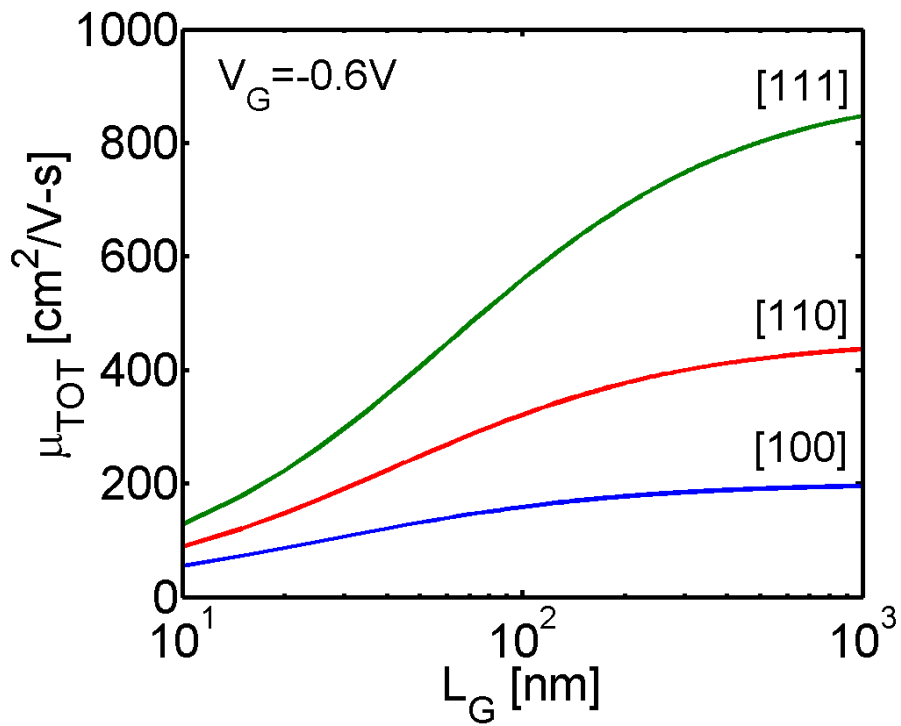

Figure 7 caption:

Total hole "apparent" mobility for NWs of D=12nm in the [100] (blue), [110] (red), and [111] (green) transport orientations versus channel length $L_G$ at $V_G$=-0.6V.

Figure 8:

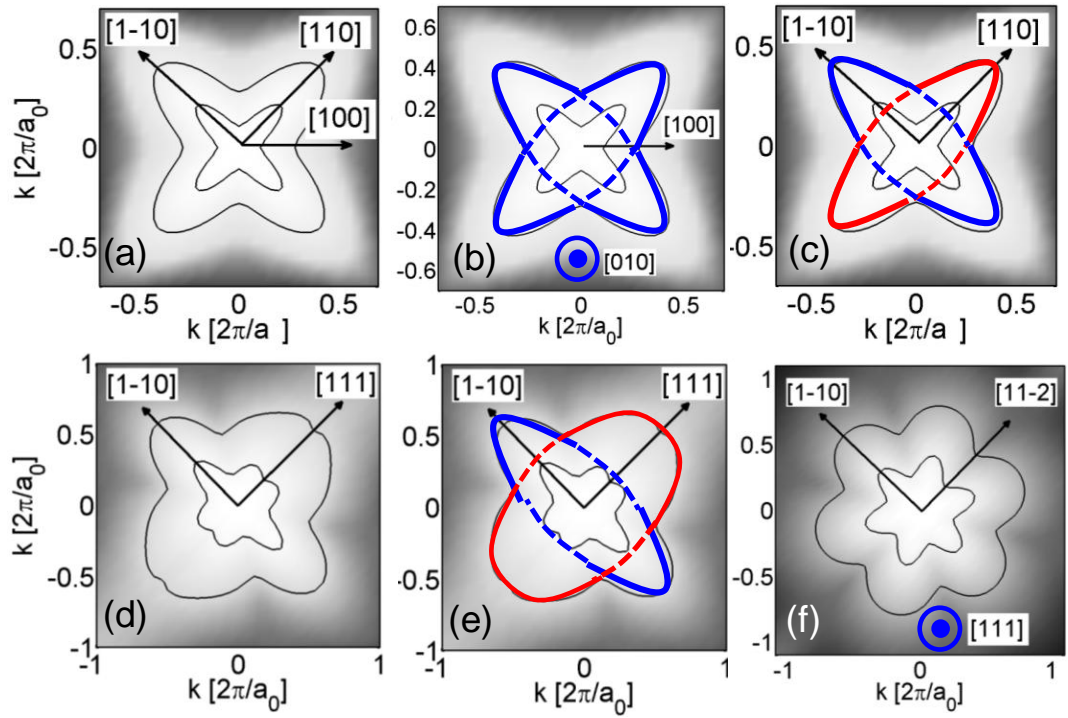

Figure 8 caption:

Energy surfaces of the heavy-hole valence band of Si. (a-c) (100) equivalent surface. (b) For the [100] NW, the symmetric [010] NW confinement affects all four "wings" equivalently. (c) For the [110] NWs under electrostatic confinement, the ground state is mostly formed from "wings" in the [1-10] direction (blue). (d-e) The (112) surface. (e) For the [111] NWs under electrostatic confinement, the ground state is mostly formed from "wings" perpendicular to [111], i.e. in the [1-10] direction (blue). (f) The (111) surface. All directions contribute to confinement almost similarly.